\documentclass[prd,superscriptaddress,amsfonts,amssymb,amsmath,showpacs]{revtex4-2}
\usepackage{bm}
\usepackage{amsfonts}
\usepackage{latexsym}
\usepackage[latin1]{inputenc}
\usepackage{graphicx}
\usepackage{amsmath}
\usepackage{palatino}
\usepackage{mathpazo}
\usepackage{textcomp}
\usepackage{float}
\usepackage{booktabs}
\usepackage{dcolumn}
\usepackage{ragged2e}
\usepackage{hyperref}
\hypersetup{colorlinks,citecolor=green}
\hypersetup{colorlinks=true,linkcolor=blue,filecolor=blue,    urlcolor=purple}
\usepackage{amsmath}
\usepackage{xcolor}
\usepackage{orcidlink}
\usepackage[caption=false]{subfig}
\usepackage{commath}
\def\jnl@style{\it}
\def\aaref@jnl#1{{\jnl@style#1}}

\def\aaref@jnl#1{{\jnl@style#1}}

\def\aj{\aaref@jnl{AJ}}                   
\def\apj{\aaref@jnl{ApJ}}                 
\def\apjl{\aaref@jnl{ApJ}}                
\def\apjs{\aaref@jnl{ApJS}}               
\def\apss{\aaref@jnl{Ap\&SS}}             
\def\aap{\aaref@jnl{A\&A}}                
\def\aapr{\aaref@jnl{A\&A~Rev.}}          
\def\aaps{\aaref@jnl{A\&AS}}              
\def\mnras{\aaref@jnl{Mon.~Not.~Roy.~Astron.~Soc.}}             
\def\prd{\aaref@jnl{Phys.~Rev.~D}}        
\def\plb{\aaref@jnl{Phys.~Lett.~B}}        
\def\prc{\aaref@jnl{Phys.~Rev.~C}}  
\def\prl{\aaref@jnl{Phys.~Rev.~Lett.}}    
\def\qjras{\aaref@jnl{QJRAS}}             
\def\skytel{\aaref@jnl{S\&T}}             
\def\ssr{\aaref@jnl{Space~Sci.~Rev.}}     
\def\zap{\aaref@jnl{ZAp}}                 
\def\nat{\aaref@jnl{Nature}}              
\def\aplett{\aaref@jnl{Astrophys.~Lett.}} 
\def\apspr{\aaref@jnl{Astrophys.~Space~Phys.~Res.}} 
\def\physrep{\aaref@jnl{Phys.~Rep.}}      
\def\physscr{\aaref@jnl{Phys.~Scr}}       
\def\commat{\aaref@jnl{Comm.~Math.~Phys.}}              
\def\science{\aaref@jnl{Science}}               
\def\cqg{\aaref@jnl{Classical Quant.~Grav.}}            
\def\jpcs{\aaref@jnl{JPCS}}                                     
\def\ijmpd{\aaref@jnl{Int.~J.~Mod.~Phys.~D}}                    
\def\grg{\aaref@jnl{Gen.~Relat.~Gravit.}}               
\def\rpp{\aaref@jnl{Rep.~Prog.~Phys.}}          
\def\npa{\aaref@jnl{Nucl.~Phys.~A}}        
\def\lrr{\aaref@jnl{Living Rev.~Rel.}}                   
\def\jcap{\aaref@jnl{J.~Cosmology Astropart.~Phys.}}    
\def\rmp{\aaref@jnl{Rev.~Mod.~Phys.}}   
\def\epjc{\aaref@jnl{Eur.~Phys.~J.~C}}

\allowdisplaybreaks[1]

\begin{document}

\color{black}       

\title{Relativistic structure of charged quark stars in energy-momentum squared gravity  }

\author{Juan M. Z. Pretel}
 \email{juanzarate@cbpf.br}
 \affiliation{
 Centro Brasileiro de Pesquisas F{\'i}sicas, Rua Dr.~Xavier Sigaud, 150 URCA, Rio de Janeiro CEP 22290-180, RJ, Brazil
}

\author{Takol Tangphati}  
\email{takoltang@gmail.com}
\affiliation{
School of Science, Walailak University, Thasala, \\Nakhon Si Thammarat, 80160, Thailand
}

\author{Ayan Banerjee} 
\email{ayanbanerjeemath@gmail.com}
\affiliation{Astrophysics and Cosmology Research Unit, School of Mathematics, Statistics and Computer Science, University of KwaZulu--Natal, Private Bag X54001, Durban 4000, South Africa}


\date{\today}

\begin{abstract}
Within the context of energy-momentum squared gravity (EMSG), where non-linear matter contributions appear in the gravitational action, we derive the modified TOV equations describing the hydrostatic equilibrium of charged compact stars. We adopt two different choices for the matter Lagrangian density ($\mathcal{L}_m= p$ versus $\mathcal{L}_m= -\rho$) and investigate the impact of each one on stellar structure. Furthermore, considering a charge profile where the electric charge density $\rho_{\rm ch}$ is proportional to the standard energy density $\rho$, we solve numerically the stellar structure equations in order to obtain the mass-radius diagrams for the MIT bag model equation of state (EoS). For $\mathcal{L}_m= p$ and given a specific value of $\beta$ (including the uncharged case when $\beta= 0$), the maximum-mass values increase (decrease) substantially as the gravity model parameter $\alpha$ becomes more negative (positive). However, for uncharged configurations and considering $\mathcal{L}_m= -\rho$, our numerical results reveal that when we increase $\alpha$ (from a negative value) the maximum mass first increases and after reaching a maximum value it starts to decrease. Remarkably, this makes it a less trivial behavior than that caused by the first choice when we take into account the presence of electric charge ($\beta \neq 0$).

\end{abstract}

\maketitle

\section{Introduction}

Over the past few decades, the investigation of the properties of compact astrophysical objects, such as black holes, neutron stars (NSs) and white dwarfs, has been of great interest in trying to understand the physics of strongly interacting matter under extreme conditions. Specifically, NSs are of great interest in many scientific areas, like nuclear physics, quantum chromodynamics (QCD), particle astrophysics, general relativity (GR) and modified gravity theories \cite{Glendenning2000, Lattimer2016, Olmo2020}. Furthermore, information about the internal composition and evolution of a NS is a challenging topic that remains an open and tantalizing question.

It is also interesting to note that new accumulating data from X-ray astronomy satellites provide constraints on the true nature of ultra-dense compact objects, where the density can reach several times higher than nuclear saturation density at the center. Nevertheless, the state of matter at these extremely high densities is still quite hazy to us \cite{Weber:2004kj}. As a consequence, several researchers expect a rich phase diagram with exotic forms of matter, in which quarks are the relevant degrees of freedom, could be formed. According to the ``strange quark matter'' hypothesis this form of matter might be the true ground state of strongly interacting matter. If this hypothesis is correct, the stellar core could then consist of quark matter, thus giving rise to the so-called quark stars (QSs). In the original model there are three fundamental quarks nicknamed up ($u$), down ($d$) and strange ($s$) quarks. The possible existence of QSs was first predicted by Gell-Mann and Zweig \cite{GellMann:1964nj, Zweig:1964jf} arguing that some compact stars are made up not of neutrons but strange quark matter (SQM).

On the other hand, the recent progress by many independent observations has imposed strong constraints on the ultradense-matter EoS, in addition to testing the presence of deconfined quark matter in massive NSs \cite{Annala:2021gom}. On this basis, theoretical models have been proposed to study the characteristics of the different phases of quark matter. Among them, the MIT bag model is the simplest alternative EoS \cite{Farhi:1984qu} in which the hadrons are considered as the bubble of free quarks confined within a bag. Many attempts have been made to describe compact stars using this EoS in strong gravitational fields. For example, uncharged \cite{Thirukkanesh:2014dja} and charged \cite{Komathiraj:2007cw, Sunzu:2014wva, Maharaj:2014vva} exact solutions have been developed using the MIT bag model EoS. An analysis of the hydrostatic equilibrium and stability against radial perturbations of charged QSs was performed in Ref.~\cite{Arbanil:2015uoa}, and more recently for a unified interacting quark matter EoS \cite{Zhang2021}. In addition, the possible existence of stable QSs in modified gravity theories has been investigated in Refs.~\cite{Astashenok:2014dja, Jimenez:2021wik, Pretel:2022plg, Banerjee:2023tmt}, particularly focusing on the maximum-mass values from the mass-radius diagrams.

Interestingly, compact astrophysical objects are an excellent testbed to look for possible modifications of Einstein's theory and their observational signatures \cite{Psaltis:2008bb}. Beside specifying any EoS for compact stars, we should also try to preclude those modified theories of gravity (MTG) that are unable to pass through observational and experimental support. Assuming this point of view, we are going to address the validity of MTG through the strong-gravity regime. It is therefore our goal to investigate the basic properties of QSs in MTG, whereas most studies have been performed within the context of conventional Einstein gravity. Of course, another important purpose of this analysis is to show the difference between GR and MTG results that might become prominent in the strong-field regime.

Here we consider a particular theory of modified gravity by adding a nonlinear $T_{\mu\nu}T^{\mu\nu}$ term to the generic action, where $T_{\mu \nu}$ is the energy-momentum tensor. This is an extension of the $f(R)$ gravity theory that includes new types of contributions to the right-hand side of the Einstein field equations, instead of searching for new material ingredients (see some other similar types of theories in Refs.~\cite{Harko:2010mv, Harko:2011kv}). Hence, such a choice for the corresponding Lagrangian leads to $f(R,T_{\mu\nu}T^{\mu\nu})$ gravity theory, which was originally studied in \cite{Katirci:2013okf}. These matter stress generalizations modify the gravitational field equations on the right-hand side and are expected to affect the cosmological evolution at high and low density regimes \cite{Roshan:2016mbt, Akarsu:2017ohj, Board:2017ign}. Modifications of this nature give rise to a particular gravity theory in the form $f(R,\textbf{T}^2) = R+ \alpha\textbf{T}^2$, dubbed as ``energy-momentum squared gravity'' (EMSG) \cite{Roshan:2016mbt}. Notwithstanding, similar to other generalized gravity models \cite{Harko:2011kv, Harko2008PLB, Haghani2021}, EMSG theory does not satisfy the conservation of energy-momentum tensor implying the presence of an extra force which acts as a non-geodesic motion of particles \cite{Nazari2020PRD, Sharif2021EPJP}. In fact, it has been argued that the lack of energy-momentum conservation opens the possibility of a gravitationally induced particle production due to the nonminimal curvature-matter coupling \cite{Harko2015}. An important consequence of adding higher-order matter terms in the usual theory of GR is that they lead to field equations of similar form to several particular theories with different fundamental bases, including bulk viscous cosmology, loop quantum gravity, $k$-essence, and brane-world cosmologies \cite{Board:2017ign}. Through a post-Minkowskian approximation and using binary-pulsar observations, the gravitational energy flux in the framework of EMSG was recently analyzed by Nazari \textit{et al.}~\cite{Nazari2022}. Moreover, the light deflection and Shapiro time delay in the weak-field limit of EMSG were investigated in Ref.~\cite{Nazari104026}, where a new EMSG potential affects the trajectory of photons.

A more general form has also been proposed and characterized by $f(R,\textbf{T}^2) = R+ \alpha (\textbf{T}^2)^n$, dubbed as energy-momentum powered gravity (EMPG), and which describes an evolution of the Universe similar to that in the 
$\Lambda$CDM model \cite{Akarsu:2017ohj}. It was further discussed that EMPG can be unified with the Starobinsky model in order to describe the complete history of the Universe including the inflationary era. It is important to note that the so-called EMSG actually is a special case of EMPG. We should also mention that the literature offers another model of $f(R,T_{\mu\nu}T^{\mu\nu})$ gravity, which is called energy-momentum log gravity (EMLG) \cite{Akarsu:2019ygx} and it was used for an extension of the standard $\Lambda$CDM model to study viable cosmologies. With respect to the EMSG case, the cosmological solutions have been studied in Refs.~\cite{Akarsu:2018drb, Faria:2019ejh, Barbar:2019rfn, Akarsu:2020vii}. Furthermore, a dynamical system analysis of generalized EMSG was carried out by Bahamonde and collaborators \cite{Bahamonde:2019urw}, while the thermodynamic properties of the Universe in the background of the generalized EMSG theory were explored in Ref.~\cite{Rudra2021}.

Besides the cosmological aspect, the study of compact stars in EMSG would be interesting to check the viability of this theory in the strong-gravity regime. In particular, a comprehensive analysis about NS mass-radius relations for four different realistic EoS has been carried out in Ref. \cite{Akarsu:2018zxl}. In addition to these studies, compact stars have been studied using various EoSs in \cite{Nari:2018aqs, Sharif:2022zmw, Sharif:2022prm} satisfying some observational measurements. In the same gravitational framework, QSs have been studied as well (see, e.g.~Refs.~\cite{Singh:2020bdv, Tangphati:2022acb} and references therein).  In the present work we are interested in studying the possible existence of charged QSs in EMSG based on the phenomenological MIT bag model EoS. In such a system the amount of electric charge contained in a compact star may be more stable than normal star, and also prevent the gravitational collapse. Bekenstein, in 1971 \cite{Bekenstein}, generalized the TOV equations of hydrostatic equilibrium to the charged case and discussed their applicability. In pure Einstein gravity, the presence of the electric charge in compact objects was discussed by several authors \cite{Negreiros:2009fd, Ray:2003gt, Panotopoulos:2021cxu, Lemos:2014lza, Arbanil:2017huq, Ivanov:2002jy}. From the above handful discussion we are motivated to study the equilibrium structure of charged QSs within the framework of EMSG focusing on the determination of their maximum masses. We will investigate the impact of both the gravity model parameter $\alpha$ and charge parameter $\beta$ on the most basic properties of charged QSs. More precisely, we will give a brief overview on the derivation of the modified TOV equations depending on the different choices of the matter Lagrangian densities $\mathcal{L}_m= p$ and $\mathcal{L}_m= -\rho$, separately.

This work is organized as follows: The EMSG theory is briefly presented and we provide the modified field equations for Maxwell-EMSG theory in Sect.~\ref{sec2}. In the same section we derive the associated stellar structure equations that describe the star interior assuming two different choices for the matter Lagrangian density. In Sect.~\ref{sec3}, we present the quark matter EoS and the charge profile which will be necessary to describe charged QSs. Section \ref{sec4} is devoted to the numerical results for uncharged and charged QSs, and we report their global properties in terms of the free parameters. We conclude in Sect.~\ref{sec5} with a summary of the key results.


\section{Energy-momentum squared gravity}\label{sec2}

\subsection{Field equations}

In order to investigate the hydrostatic equilibrium state of charged compact stars in EMSG, here we briefly summarize the gravitational background taking the Maxwell contributions into account. Under this perspective, we first introduce the EMSG field equations in the presence of electromagnetic field. Namely, the Einstein-Hilbert action is modified as
\begin{equation}\label{action}
    S = \int d^4x\sqrt{-g} \left[ \frac{R}{16\pi} + \alpha T_{\mu\nu}T^{\mu\nu} + \mathcal{L}_m + \mathcal{L}_e \right] ,
\end{equation}
where $g$ is the determinant of the metric, $R$ is the Ricci scalar, $T_{\mu\nu}$ is the energy-momentum tensor for the standard matter distribution associated with the matter Lagrangian density $\mathcal{L}_m$, and $\mathcal{L}_e$ denotes the Lagrangian for the electromagnetic field. The constant $\alpha$ is a free parameter of EMSG modification with respect to the conventional GR theory, and which is given in $\rm m^{-2}$ units within a geometric unit system.

Varying the action (\ref{action}) with respect to the inverse metric, one obtains the field equations
\begin{equation}\label{FieldEq}
    G_{\mu\nu} = 8\pi \left( T_{\mu\nu} + \mathcal{E}_{\mu\nu} \right) + 8\pi\alpha \left( g_{\mu\nu} T_{\sigma\rho}T^{\sigma\rho} - 2\Theta_{\mu\nu} \right) ,
\end{equation}
where $G_{\mu\nu}$ is the usual Einstein tensor, $\mathcal{E}_{\mu\nu}$ is the electromagnetic energy-momentum tensor, and the new tensor $\Theta_{\mu\nu}$ is defined as 
\begin{equation}\label{ThetaEq}
    \Theta_{\mu\nu} \equiv T^{\sigma\rho}\frac{\delta T_{\sigma\rho}}{\delta g^{\mu\nu}} + T_{\sigma\rho}\frac{\delta T^{\sigma\rho}}{\delta g^{\mu\nu}} = 2T_\mu^\sigma T_{\nu\sigma} - 2\mathcal{L}_m\left[ T_{\mu\nu} - \frac{1}{2}g_{\mu\nu}T \right] - TT_{\mu\nu} - 4T^{\sigma\rho} \frac{\partial^2 \mathcal{L}_m}{\partial g^{\mu\nu} \partial g^{\sigma\rho}}, 
\end{equation}
with $T$ being the trace of the energy-momentum tensor $T_{\mu\nu}$. Besides, the relation between $T_{\mu\nu}$ and $\mathcal{L}_m$ is given by
\begin{equation}
    T_{\mu \nu} = \frac{-2}{\sqrt{-g}} \frac{\delta(\sqrt{-g}\mathcal{L}_m)}{\delta g^{\mu\nu}} = g_{\mu\nu}\mathcal{L}_m - 2\frac{\partial\mathcal{L}_m}{\partial g^{\mu\nu}} .
\end{equation}

Following a similar procedure to the one presented by Ray \textit{et al.}~\cite{Ray:2003gt} in the pure GR scenario, we consider that the stellar fluid is made of an isotropic perfect fluid plus an electromagnetic field. Consequently, the matter and electromagnetic energy-momentum tensors have the following form, respectively,
\begin{equation}\label{MatterEMT}
    T_{\mu\nu} = (\rho+ p)u_\mu u_\nu + pg_{\mu\nu} ,
\end{equation}
and 
\begin{equation}\label{ElectEMT}
    \mathcal{E}_{\mu\nu} = \frac{1}{4\pi} \left[ F_{\mu\lambda}g^{\alpha\lambda}F_{\nu\alpha} - \frac{1}{4}g_{\mu\nu}F_{\lambda\sigma}F^{\lambda\sigma} \right] ,
\end{equation}
where $\rho$ is the energy density, $p$ is the pressure and $u^\mu$ is the four-velocity of the fluid. The antisymmetric electromagnetic field strength tensor $F_{\mu\nu}$ can be defined in terms of the electromagnetic four-potential $A_\mu$ through $F_{\mu\nu} = \nabla_\mu A_\nu - \nabla_\nu A_\mu$, where $\nabla_\mu$ represents the covariant derivative. Such strength tensor must satisfy the Maxwell equations: $\nabla_\mu F^{\mu\nu} = -4\pi j^\nu$ and $\nabla_{[\sigma}F_{\mu\nu]} =0$, or alternatively
\begin{align}
    &\frac{1}{\sqrt{-g}} \frac{\partial}{\partial x^\mu}\left( \sqrt{-g}F^{\mu\nu} \right) = -4\pi j^\nu ,  \label{MaxwEq1}  \\
    &\nabla_\sigma F_{\mu\nu} + \nabla_\mu F_{\nu\sigma} + \nabla_\nu F_{\sigma\mu} = 0 ,  \label{MaxwEq2}
\end{align}
where $j^\mu = \rho_{\rm ch}u^\mu$ is the four-current density and $\rho_{\rm ch}$ is the electric charge density.

To derive the modified TOV equations that describe the internal structure of a compact star in EMSG, we choose the static spherically symmetric line element in the form
\begin{equation}\label{MetricEq}
    ds^2 = g_{\mu\nu}dx^\mu dx^\nu = -e^{2\psi}dt^2 + e^{2\lambda}dr^2 + r^2(d\theta^2 + \sin^2\theta d\phi^2) ,
\end{equation}
where $x^\mu = (t,r,\theta,\phi)$ is the 4-position vector, and the metric functions $\psi$ and $\lambda$ depend only on the radial coordinate $r$. This entails that $\sqrt{-g}= e^{\psi+ \lambda}r^2\sin\theta$ and $u^\mu = e^{-\psi}\delta_0^\mu$. In addition, assuming that the electromagnetic field is generated solely due to the electric charge, the only non-vanishing component of the strength tensor is $F^{01} = -F^{10}$. As a consequence, the Maxwell equation (\ref{MaxwEq1}) leads to
\begin{equation}
    F^{01} = \frac{q(r)}{r^2}e^{-\psi - \lambda} , 
\end{equation}
where $q(r)$ represents a charge function within a sphere of radius $r$ in the charged fluid, given by
\begin{equation}\label{ChargeEq}
    q(r) = 4\pi\int _0^r \bar{r}^2\rho_{\rm ch}(\bar{r})e^{\lambda(\bar{r})}d\bar{r} .
\end{equation}

Bearing in mind that $\nabla^\mu G_{\mu\nu}= 0$, the covariant divergence of the field equations (\ref{FieldEq}) leads to the non-conservation of energy-momentum, namely
\begin{equation}
    \nabla^\mu T_{\mu \nu} = -\nabla^\mu\mathcal{E}_{\mu\nu} - \alpha g_{\mu\nu} \nabla^\mu (T_{\sigma\rho} T^{\sigma\rho}) + 2\alpha \nabla^\mu \Theta_{\mu\nu} ,
\end{equation}
which, in view of Eqs.~(\ref{MatterEMT}) and (\ref{ElectEMT}), can also be written as
\begin{equation}\label{NonConservEq}
    \nabla^\mu T_{\mu \nu} = - j^\lambda F_{\lambda\nu} - \alpha \nabla_\nu \left(\rho^2+ 3p^2\right) + 2\alpha \nabla^\mu \Theta_{\mu\nu} .
\end{equation}
Note that the expression for the tensor $\Theta_{\mu\nu}$ in Eq.~(\ref{ThetaEq}) depends on the particular choice of the matter Lagrangian density $\mathcal{L}_m$. Since this choice is not unique, in the following we will derive the stellar structure equations for the two cases provided in the literature \cite{Bertolami2008, Faraoni2009}, i.e.~$\mathcal{L}_m = p$ and $\mathcal{L}_m = -\rho$. In fact, it has been shown that for a perfect fluid that does not couple explicitly to the curvature, the two Lagrangian densities are perfectly equivalent.

\subsection{Modified TOV equations for $\mathcal{L}_m = p$}

When we adopt the choice $\mathcal{L}_m = p$, we obtain $\Theta_{\mu\nu} = -(\rho^2+ 4\rho p + 3p^2)u_\mu u_\nu$, so that Eqs.~(\ref{FieldEq}) and (\ref{NonConservEq}) assume the form
\begin{align}
    &G_{\mu\nu} = 8\pi\rho \left[ \left( 1+ \frac{p}{\rho} \right)u_\mu u_\nu + \frac{p}{\rho}g_{\mu\nu} \right] + 8\pi\mathcal{E}_{\mu\nu} + 8\pi\alpha\rho^2 \left[ \left( 1+ 3\frac{p^2}{\rho^2} \right)g_{\mu\nu} + 2\left( 1+ 4\frac{p}{\rho}+ 3\frac{p^2}{\rho^2} \right)u_\mu u_\nu \right] ,  \label{FieldEqLmp}  \\
    &\nabla^\mu T_{\mu \nu} = - j^\lambda F_{\lambda\nu} - \alpha \partial_\nu \left(\rho^2+ 3p^2\right) - 2\alpha\left( \rho^2+ 4\rho p+ 3p^2 \right)\Gamma_{0\nu}^0 .  \label{NonConsevEqLmp}
\end{align}
Through the metric (\ref{MetricEq}), the $00$ and $11$ components of the field equations (\ref{FieldEqLmp}) are given explicitly by 
\begin{align}
    &\frac{1}{r^2}\frac{d}{dr}\left( re^{-2\lambda} \right) - \frac{1}{r^2} = -8\pi\left( \rho + \frac{q^2}{8\pi r^4} \right) - 8\pi\alpha\rho^2\left( 1+ 8\frac{p}{\rho}+ 3\frac{p^2}{\rho^2} \right) ,  \label{Eq16} \\
    &e^{-2\lambda}\left( \frac{2}{r}\psi'+ \frac{1}{r^2} \right) - \frac{1}{r^2} = 8\pi \left( p - \frac{q^2}{8\pi r^4} \right) + 8\pi\alpha\rho^2\left( 1+ 3\frac{p^2}{\rho^2} \right) ,  \label{Eq17}
\end{align}
where the prime stands for differentiation with respect to $r$. In the absence of electric charge (i.e., $q= 0$), such equations are reduced to those obtained in Ref.~\cite{Akarsu:2018zxl}. Furthermore, for the index $\nu= 1$, the non-conservation equation (\ref{NonConsevEqLmp}) leads to 
\begin{equation}\label{pPrimeEq}
    p' = -\frac{\rho +p}{1+ 6\alpha p}\left[ 1+ 2\alpha\rho\left( 1+ 3\frac{p}{\rho} \right) \right]\psi' + \frac{qq'}{4\pi r^4(1+ 6\alpha p)} - \frac{2\alpha\rho\rho'}{1+ 6\alpha p} . 
\end{equation}

In order to write the stellar structure equations in a more familiar form as in the pure GR case, it is convenient to introduce a mass parameter that allows us to calculate the gravitational mass of a charged fluid sphere within the context of EMSG model. With this in mind, integrating Eq.~(\ref{Eq16}) we arrive at
\begin{equation}
    e^{-2\lambda} = 1- \frac{2}{r}\left\lbrace 4\pi\int r^2\rho dr + \frac{1}{2}\int \frac{q^2}{r^2}dr + 4\pi\alpha\int r^2\rho^2\left( 1+ 8\frac{p}{\rho} + 3\frac{p^2}{\rho^2} \right)dr \right\rbrace ,
\end{equation}
or alternatively
\begin{equation}\label{ExpLambda}
    e^{-2\lambda} = 1 - \frac{2m}{r} + \frac{q^2}{r^2} ,
\end{equation}
where the mass function $m(r)$ can be understood as the total gravitational mass contained within a charged sphere of radius $r$, 
\begin{equation}\label{MassFunc}
    m = 4\pi\int r^2\rho dr + \int\frac{qq'}{r}dr + 4\pi\alpha\int r^2\rho^2\left( 1+ 8\frac{p}{\rho} + 3\frac{p^2}{\rho^2} \right)dr .
\end{equation}

In other words, the total mass of the star is the contribution of three terms: energy density of conventional matter in the first integral, electric charge via $q(r)$, and EMSG modification through the third integral. For the particular case $\alpha= 0$, we retrieve the standard mass function of a charged perfect fluid in Einstein gravity \cite{Negreiros:2009fd, Arbanil2013}. Moreover, in the uncharged situation, the above equation reduces to the expression given by Akarsu and collaborators \cite{Akarsu:2018zxl}. 

Using Eq.~(\ref{ExpLambda}), the 11-component of the field equations (\ref{Eq17}) becomes 
\begin{equation}\label{PsiPrimeEq}
    \psi' = \left[ \frac{m}{r^2} + 4\pi rp - \frac{q^2}{r^3} + 4\pi\alpha r\rho^2\left( 1+ 3\frac{p^2}{\rho^2} \right) \right]e^{2\lambda} . 
\end{equation}
Consequently, from Eqs.~(\ref{ChargeEq}), (\ref{pPrimeEq}), (\ref{MassFunc}) and (\ref{PsiPrimeEq}), we obtain the following modified TOV equations 
\begin{align}
    \frac{dq}{dr} =&\ 4\pi r^2\rho_{\rm ch} \left( 1- \frac{2m}{r}+ \frac{q^2}{r^2} \right)^{-1/2} ,  \label{TOV1Lmp}  \\  
    \frac{dm}{dr} =&\ 4\pi r^2\rho + \frac{qq'}{r} + 4\pi\alpha r^2\rho^2\left( 1+ 8\frac{p}{\rho}+ 3\frac{p^2}{\rho^2} \right),  \label{TOV2Lmp}  \\ 
    \frac{dp}{dr} =& -\frac{\rho+ p}{1+ 6\alpha p}\left[ 1+ 2\alpha\rho\left( 1+ 3\frac{p}{\rho} \right) \right] \left[ \frac{m}{r^2} + 4\pi rp- \frac{q^2}{r^3} + 4\pi\alpha r\rho^2\left( 1+ 3\frac{p^2}{\rho^2} \right) \right]\left( 1- \frac{2m}{r}+ \frac{q^2}{r^2} \right)^{-1}  \nonumber  \\
    &+ \frac{qq'}{4\pi r^4(1+ 6\alpha p)} - \frac{2\alpha\rho\rho'}{1+ 6\alpha p} ,  \label{TOV3Lmp}  \\
    \frac{d\psi}{dr} =& -\frac{1}{\rho+ p}\left[ (1+ 6\alpha p)p' + 2\alpha\rho\rho' - \frac{qq'}{4\pi r^4} \right]\left[ 1+ 2\alpha\rho\left( 1+ 3\frac{p}{\rho} \right) \right]^{-1} ,  \label{TOV4Lmp}
\end{align}
which describe the hydrostatic equilibrium of a charged compact star within the framework of energy-momentum squared gravity. By setting the parameter $\alpha =0$, the conventional TOV equations are recovered. As expected, when the electric charge vanishes (that is, $\rho_{\rm ch}= 0$ and thus $q= 0$), the system of differential equations (\ref{TOV1Lmp})-(\ref{TOV4Lmp}) reduces to the modified TOV equations obtained in Ref.~\cite{Akarsu:2018zxl}.

Similar to the general relativistic case, the above modified TOV equations are solved numerically for a specific EoS relating to the pressure and energy density of the fluid, namely $p= p(\rho)$. Furthermore, this set of equations must be supplied with an electrical charge profile in order to close the system. If we assume that $\rho_{\rm ch} = \rho_{\rm ch}(\rho)$, we will have four unknown variables to be determined. To guarantee regularity of spacetime geometry, we have to specify the following boundary conditions at the stellar origin
\begin{align}\label{BC1}
    q(0) &= 0,   &   m(0) &= 0 ,   &   \rho(0) &= \rho_c ,
\end{align}
where $\rho_c$ is the central energy density. We integrate  Eqs.~(\ref{TOV1Lmp})-(\ref{TOV3Lmp}) from the center of the star up to the radial coordinate where pressure vanishes. In other words, the stellar surface is determined when $p(r_{\rm sur})= 0$, where $r_{\rm sur}$ denotes the radius of the charged star.

On the other hand, if we want to determine the radial behavior of the metric potential $\psi$ we will have to provide an extra boundary condition. With this in mind, the trace of the field equations \eqref{FieldEqLmp} allows us to obtain the Ricci scalar in terms of the fluid variables as $R= 8\pi(\rho- 3p)[1- 2\alpha(\rho- p)]$. In the outer region of the star (where $\rho= p =0$) we therefore have $R=0$, so that the exterior spacetime is still described by the conventional Reissner-Nordstr{\"o}m solution as in Einstein gravity. As a consequence, the continuity of the metric at the surface of the star imposes a boundary condition for the differential equation (\ref{TOV4Lmp}), namely
\begin{equation}
    \psi(r_{\rm sur}) = \frac{1}{2}\ln\left[ 1 - \frac{2M}{r_{\rm sur}} + \frac{Q^2}{r_{\rm sur}^2} \right] ,
\end{equation}
where $M \equiv m(r_{\rm sur})$ and $Q \equiv q(r_{\rm sur})$ are the total gravitational mass and total electric charge determined at the surface, respectively.

\subsection{Modified TOV equations for $\mathcal{L}_m = -\rho$}

As argued by Bertolami \textit{et al.}~\cite{Bertolami2008}, $\mathcal{L}_m =p$ is not the unique choice for the matter Lagrangian density. In that regard, we are also going to explore the other case when $\mathcal{L}_m =-\rho$. This implies that $\Theta_{\mu\nu}= (\rho^2- p^2)(u_\mu u_\nu+ g_{\mu\nu})$, and hence Eqs.~(\ref{FieldEq}) and (\ref{NonConservEq}) become 
\begin{align}
    &G_{\mu\nu} = 8\pi\rho \left[ \left( 1+ \frac{p}{\rho} \right)u_\mu u_\nu + \frac{p}{\rho}g_{\mu\nu} \right] + 8\pi\mathcal{E}_{\mu\nu} + 8\pi\alpha\rho^2 \left[ \left( 5\frac{p^2}{\rho^2}- 1 \right)g_{\mu\nu} - 2\left( 1- \frac{p^2}{\rho^2} \right)u_\mu u_\nu \right] ,  \label{FieldEqLmMinusRho}  \\
    &\nabla^\mu T_{\mu \nu} = - j^\lambda F_{\lambda\nu} + \alpha \partial_\nu \left(\rho^2- 5p^2\right) + 2\alpha\left( \rho^2- p^2 \right)\Gamma_{0\nu}^0 .  \label{NonConsevEqLmMinusRho}
\end{align}
Following a similar procedure to the previous choice, the hydrostatic equilibrium equations are now given by
\begin{align}
    \frac{dq}{dr} =&\ 4\pi r^2\rho_{\rm ch} \left( 1- \frac{2m}{r}+ \frac{q^2}{r^2} \right)^{-1/2} ,  \label{TOV1LmMinusRho}  \\  
    \frac{dm}{dr} =&\ 4\pi r^2\rho + \frac{qq'}{r} - 4\pi\alpha r^2\rho^2\left( 1+ 3\frac{p^2}{\rho^2} \right),  \label{TOV2LmMinusRho}  \\ 
    \frac{dp}{dr} =& -\frac{\rho+ p}{1+ 10\alpha p}\left[ 1- 2\alpha\rho\left( 1- \frac{p}{\rho} \right) \right] \left[ \frac{m}{r^2} + 4\pi rp- \frac{q^2}{r^3} - 4\pi\alpha r\rho^2\left( 1- 5\frac{p^2}{\rho^2} \right) \right]\left( 1- \frac{2m}{r}+ \frac{q^2}{r^2} \right)^{-1}  \nonumber  \\
    &+ \frac{qq'}{4\pi r^4(1+ 10\alpha p)} + \frac{2\alpha\rho\rho'}{1+ 10\alpha p} ,  \label{TOV3LmMinusRho}  \\
    \frac{d\psi}{dr} =& -\frac{1}{\rho+ p}\left[ (1+ 10\alpha p)p' - 2\alpha\rho\rho' - \frac{qq'}{4\pi r^4} \right]\left[ 1- 2\alpha\rho\left( 1- \frac{p}{\rho} \right) \right]^{-1} ,  \label{TOV4LmMinusRho}
\end{align}
where Eq.~(\ref{TOV1LmMinusRho}) remains the same for both choices of $\mathcal{L}_m$ because it arises from Maxwell equations and not from the field equations. Moreover, note that the boundary conditions for this system of differential equations would be the same as for the case $\mathcal{L}_m= p$. The uncharged situation for the choice $\mathcal{L}_m= -\rho$ has not yet been studied in the literature, so the purpose of the present work is to fill such gap by also adopting this choice. In the GR limit, i.e.~when $\alpha \rightarrow 0$, we recover the standard TOV equations describing a charged fluid sphere as given in Ref.~\cite{Arbanil2013}.


\section{Quark matter equation of state} \label{sec3}

Quantum Chromodynamics (QCD) that describes strong interactions between quarks and gluons is an extremely successful theory. In spite of this benchmark advance, scientists believe that some extreme physics might take place inside the core of the neutron stars (NSs). The proposal that if matter is compressed far enough beyond nuclear density then the neutrons themselves will be crushed out of existence for a more exotic forms of matter, notably quark matter. This possibility was suggested about four decades ago \cite{Witten:1984rs,Bodmer:1971we} that compact stars made of strange quark matter (SQM) either located only in the inner core or present up to their surfaces. Therefor, our focus is mainly on dense stellar objects that are generically called quark stars (QSs).

Another reason for studying SQM is that it may be considered a true ground state of strongly interacting matter consisting of almost equal numbers of up ($u$), down ($d$) and strange ($s$)   quarks with the charge neutrality maintained by the inclusion
of electrons at high density and/or temperature. In order to investigate quark stars in EMSG, we consider a conceptually simple phenomenological model, namely, the conventional MIT bag model EoS:
\begin{equation}\label{MITbagEoS}
    p = \dfrac{1}{3}(\rho - 4B) ,
\end{equation}
where the bag constant $B$ corresponds to the outwards pressure exerted on the bag, and it lies in the range $0.982B_0< B <1.525B_0$ where $B_0 = 60\, \rm MeV/fm^3$. In the present work, we will consider the particular case $B= B_0$. It is worth emphasizing that, in the framework of the MIT bag model, the influence of strong magnetic field on the structure properties of strange quark stars was investigated in Ref.~\cite{Kayanikhoo2020}. Furthermore, within the context of modified gravity theories, the MIT bag model EoS was used by several authors to study quark stars \cite{Banerjee:2023tmt, Pace2017, Salako2021, Abbas2021, Biswas2021}.

In addition to the equation of state for the dense matter involved, we need to adopt a charge profile for $\rho_{\rm ch}$. In Einstein gravity, Ray and collaborators \cite{Ray:2003gt} assumed that the charge distribution is proportional to the energy density, namely $\rho_{\rm ch}= \beta\rho$. This assumption is reasonable because a large mass can hold a large amount of electric charge, where $\beta$ is a dimensionless parameter that controls the amount of electric charge within the stellar fluid \cite{Arbanil2013}. This profile was also considered to study the effect of electric charge on the quark star structure in some theories of modified gravity such as metric $f(R)$ theories \cite{Pretel2022}, $f(R,T)$ gravity \cite{Pretel2022CPC} and 4D Einstein-Gauss-Bonnet gravity \cite{Pretel2022EPJC}.


\section{Numerical results}\label{sec4}

\subsection{Uncharged stellar configurations}

\begin{figure}[h]
    \centering
    \includegraphics[width = 8.46cm]{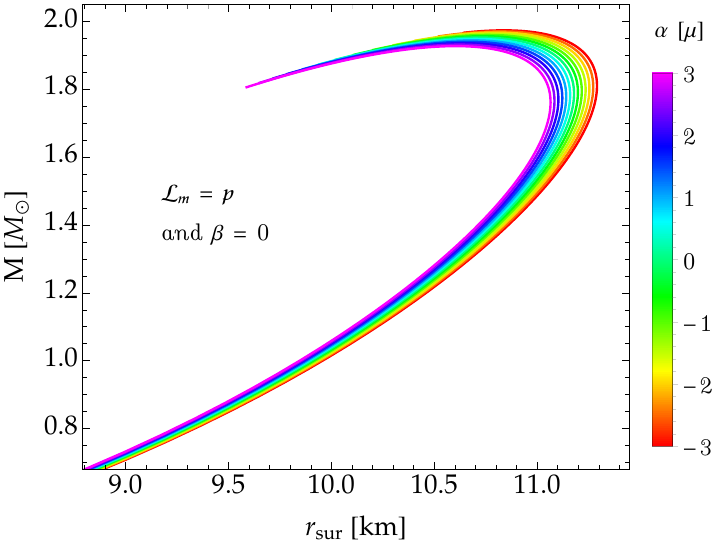}
    \includegraphics[width = 8.7cm]{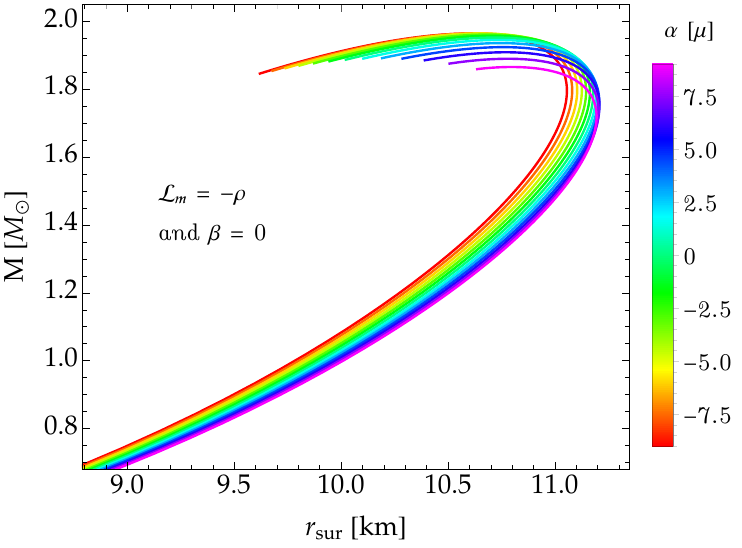}
    \caption{Mass-radius diagrams for quark stars without electric charge (i.e., when $\beta= 0$) in energy-momentum squared gravity by adopting two choices for the matter Lagrangian density: $\mathcal{L}_m =p$ (left panel) and $\mathcal{L}_m =-\rho$ (right panel). We have considered a wide range of values for the free parameter $\alpha$ (see the color scale on the right-hand side of each plot) such that appreciable changes can be visualized in the curves with respect to the pure GR case ($\alpha= 0$), where $\mu= 10^{-38}\, \rm cm^3/erg$. }
    \label{FigsUnchargedStars}
\end{figure}

Given an EoS, the numerical procedure carried out in this work is similar to that of Einstein gravity, with the difference that now we have to assume a specific value for the gravity model parameter $\alpha$. We begin our analysis considering the case of configurations without electric charge, namely, assuming $\beta= 0$ (and hence $\rho_{\rm ch}= q= 0$) in the modified TOV equations. For the particular choice $\mathcal{L}_m= p$, we numerically integrate the differential equations (\ref{TOV2Lmp}) and (\ref{TOV3Lmp}) demanding the vanishing of the mass at the origin and giving a specific value of $\alpha$ and central energy density $\rho_c$. The surface of the star is determined when the pressure vanishes, i.e., it is defined by $p(r_{\rm sur})= 0$, so that the total mass of the star is given by $M= m(r_{\rm sur})$. The left panel of Fig.~\ref{FigsUnchargedStars} shows the mass-radius relation for uncharged quark stars with MIT bag model EoS (\ref{MITbagEoS}) in EMSG for a range of central densities, where we have considered different values of $\alpha \in [-3.0, 3.0] \mu$ with $\mu= 10^{-38}\, \rm cm^3/erg$. One can see that the impact of the contribution $\alpha T_{\mu\nu}T^{\mu\nu}$ is indeed an increase (decrease) in the maximum-mass values as $\alpha$ decreases (increases) with respect to its general relativistic counterpart. A similar behavior occurs with the radius; $r_{\rm sur}$ increases with the decrease of the parameter $\alpha$.

On the other hand, for the matter Lagrangian density $\mathcal{L}_m= -\rho$, we have to solve Eqs.~(\ref{TOV2LmMinusRho})-(\ref{TOV3LmMinusRho}) with the same boundary conditions as considered for the other choice. In the right panel of Fig.~\ref{FigsUnchargedStars}, we illustrate the $M$-$r_{\rm sur}$ diagram for various values of the free parameter $\alpha \in [-9.0, 9.0] \mu$. Notice that in this case we have used larger values for $\alpha$ than in the choice $\mathcal{L}_m= p$ (see left panel) in order to obtain appreciable modifications with respect to the pure GR case. Nevertheless, such a choice leads to a peculiar behavior: When we increase $\alpha$ (from a negative value, see the red curve for $\alpha= -9.0\mu$) the maximum-mass values first increase and after reaching a maximum value they start to decrease.

\subsection{Charged stellar configurations}

\begin{figure}[h]
    \centering
    \includegraphics[width = 8.59cm]{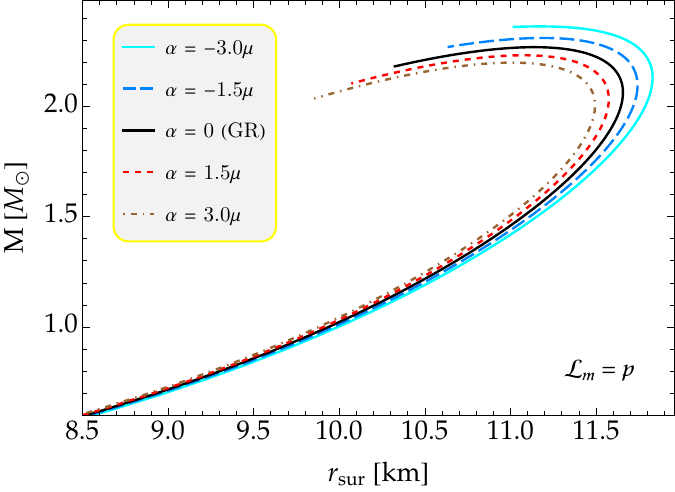}
    \includegraphics[width = 8.5cm]{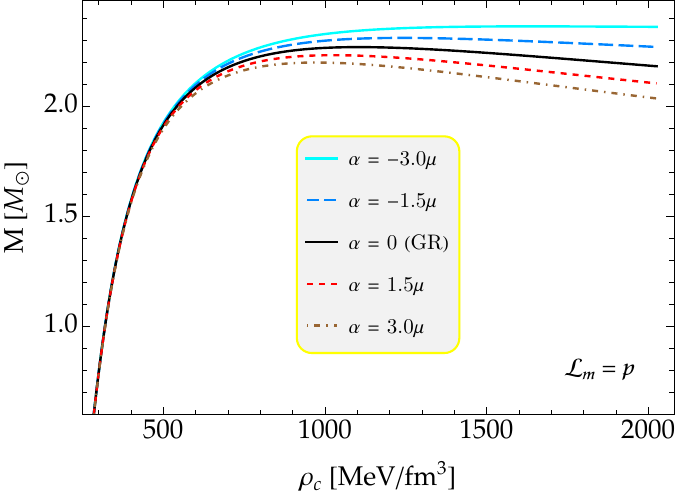}
    \includegraphics[width = 8.58cm]{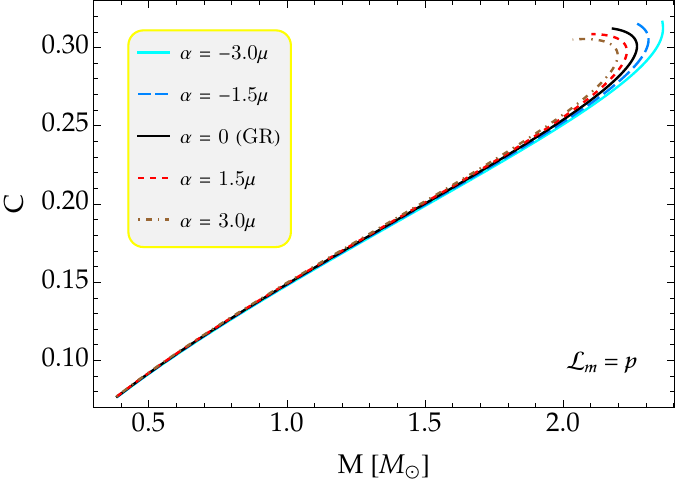}
    \includegraphics[width = 8.5cm]{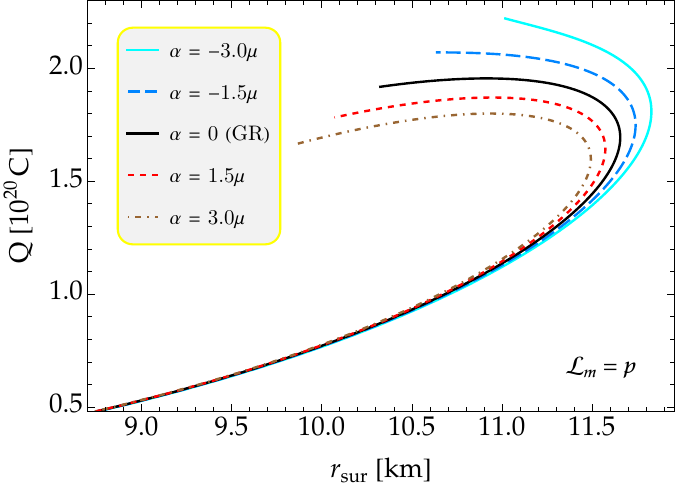}
    \caption{Mass-radius (top left), mass-central energy density (top right), compactness-mass (bottom left), and charge-radius (bottom right) relations for charged QSs with MIT bag model EoS in the EMSG theory for the matter Lagrangian density $\mathcal{L}_m= p$. Moreover, we have used the parameter set: $B = 60\, \rm MeV/fm^3$, $\beta = 0.4$, and $\alpha \in [-3.0, 3.0] \mu$. The black solid curves represent the pure general relativistic solutions in all plots.}
    \label{FigsLmp_vary_alpha}
\end{figure}

\begin{figure}
    \centering
    \includegraphics[width = 8.5cm]{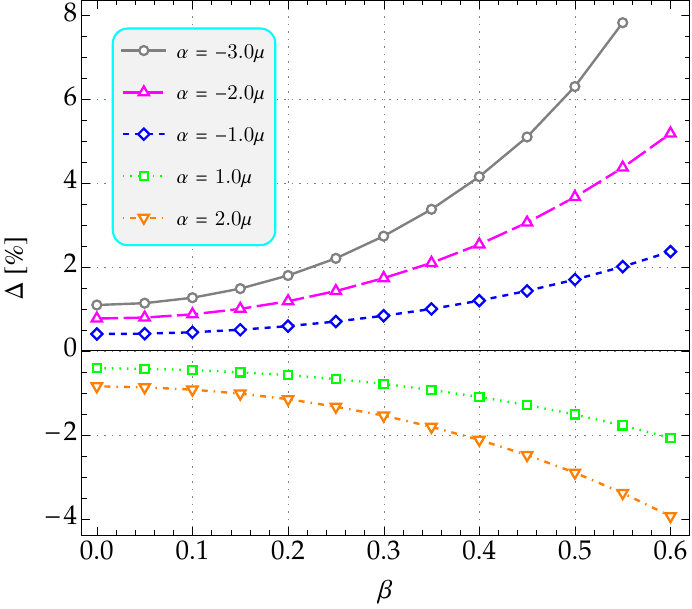}
    \includegraphics[width = 8.5cm]{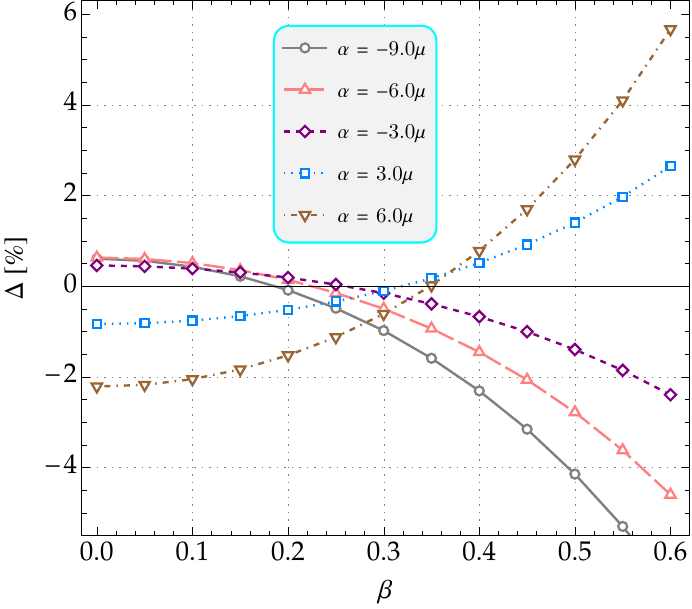}
    \caption{Relative deviation (\ref{RelativeEq}) as a function of the charge parameter $\beta$ for both matter Lagrangian densities $\mathcal{L}_m =p$ (left panel) and $\mathcal{L}_m =-\rho$ (right panel). In both panels, the uncharged case corresponds to $\beta= 0$. For $\mathcal{L}_m =p$, the maximum mass gets further and further away from its GR value as $\vert\alpha\vert$ increases. Meanwhile, for $\mathcal{L}_m =-\rho$, the maximum mass decreases (increases) and after reaching a certain value of $\beta$ it starts to increase (decrease) for positive (negative) $\alpha$. }
    \label{FigsRelDeviation}
\end{figure}

We are now going to consider QSs within the context of EMSG theory in the presence of electric charge, that is, when $\beta\neq 0$. For the case $\mathcal{L}_m= p$, we solve the modified TOV equations (\ref{TOV1Lmp})-(\ref{TOV3Lmp}) under the boundary conditions (\ref{BC1}). For a fixed value of $\beta= 0.4$ and $\alpha$ varying, in Fig.~\ref{FigsLmp_vary_alpha} we show the different macroscopic properties of charged QSs. According to the mass-radius relation (upper left plot), both the mass and radius increase as $\alpha$ becomes more negative, while the opposite occurs for positive $\alpha$. For this value of $\beta$, the relative difference between the maximum-mass values predicted by GR ($\alpha= 0$) and EMSG (with $\alpha= -3.0\mu$) is about $4.1\%$ (see the left plot of Fig.~\ref{FigsRelDeviation}). The most substantial changes for the mass and compactness due to the parameter $\alpha$ take place in the high-central-density region, while the modifications are irrelevant in the low-mass branch (see upper right and lower left plots). The lower right plot suggests that less maximum-total charge than GR is allowed for $\alpha >0$, while much more electric charge can be supported for $\alpha <0$. Remarkably, for negative $\alpha$ with sufficiently large $\vert\alpha\vert$, it is not possible to find a maximum-charge point as the curve grows indeterminately.

\begin{figure}[h]
    \centering
    \includegraphics[width = 8.58cm]{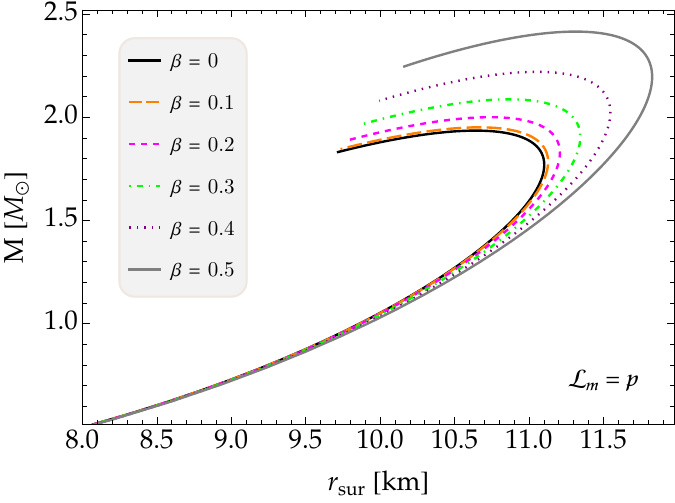}
    \includegraphics[width = 8.5cm]{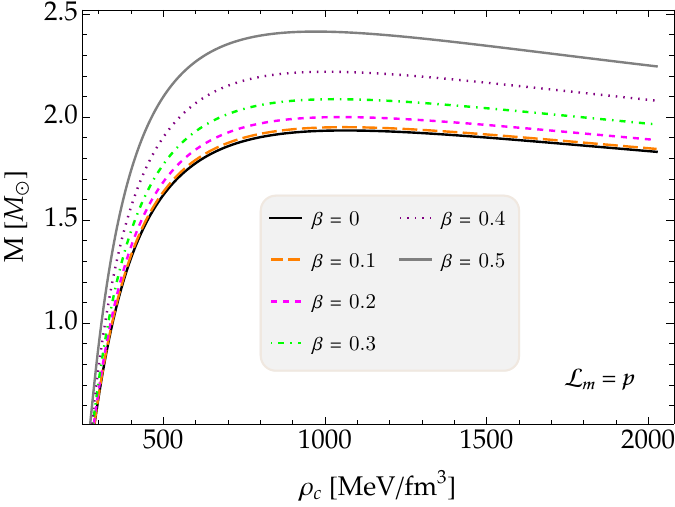}
    \includegraphics[width = 8.58cm]{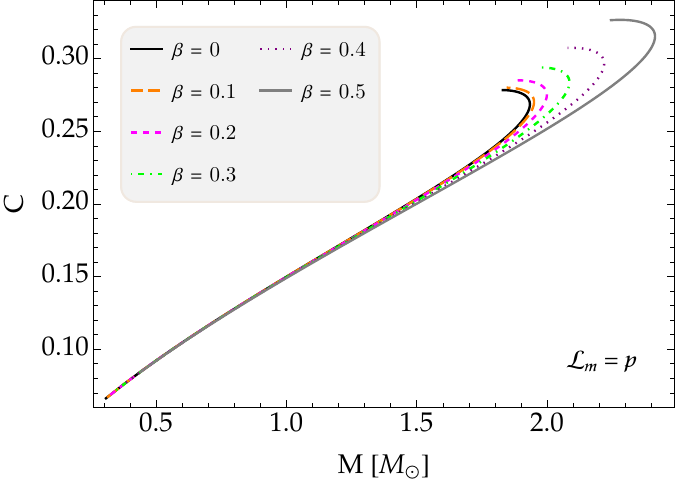}
    \includegraphics[width = 8.5cm]{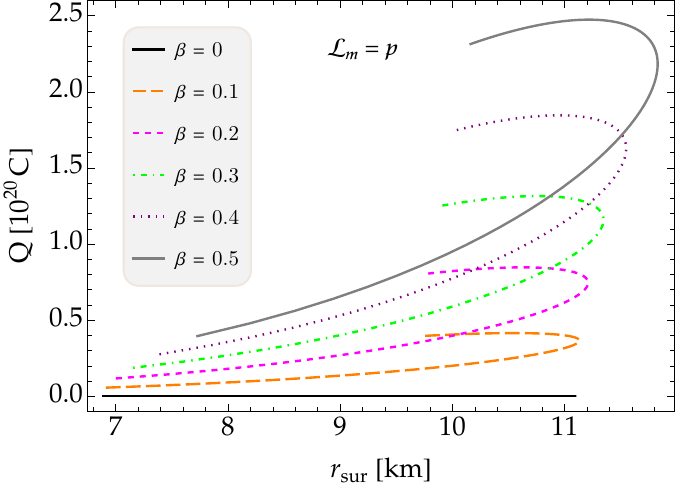}
    \caption{Macro-physical properties of charged QSs predicted by EMSG theory when $\alpha= 2.0\mu$ and $\mathcal{L}_m= p$. In addition, we have adopted $B = 60\, \rm MeV/fm^3$ for the bag constant and varied the charge parameter in the range $\beta \in [0, 0.5]$, where the particular case $\beta= 0$ represents a family of uncharged quark stars.}
    \label{FigsLmp_vary_beta}
\end{figure}

Keeping the gravity model parameter fixed $\alpha= 2.0\mu$ and varying $\beta$, we obtain the numerical results shown in Fig.~\ref{FigsLmp_vary_beta} for $\mathcal{L}_m= p$. The results are qualitatively similar to those obtained in pure Einstein gravity. Namely, the maximum-mass values increase significantly as $\beta$ increases. We also observe that the largest effect of electric charge on compactness occurs in the high-mass region. The behavior of the total charge of the relativistic sphere as a function of the radius is shown in the lower right panel. As expected, $Q$ is strongly dependent on the charge parameter $\beta$ in comparison to the plot of Fig.~\ref{FigsLmp_vary_alpha}, where the most relevant changes in the total charge due to $\alpha$ take place only in the high-charge region.

\begin{figure}[h]
    \centering
    \includegraphics[width = 8.59cm]{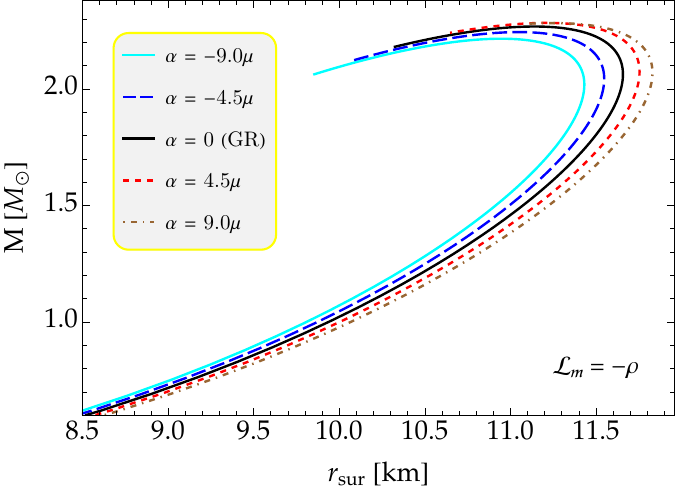}
    \includegraphics[width = 8.5cm]{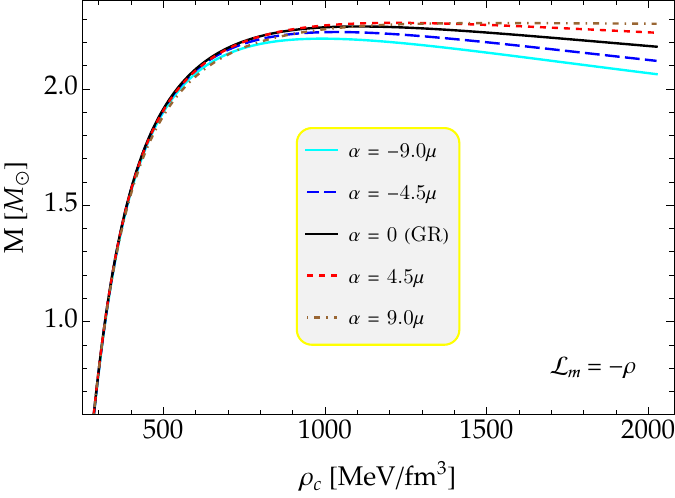}
    \includegraphics[width = 8.58cm]{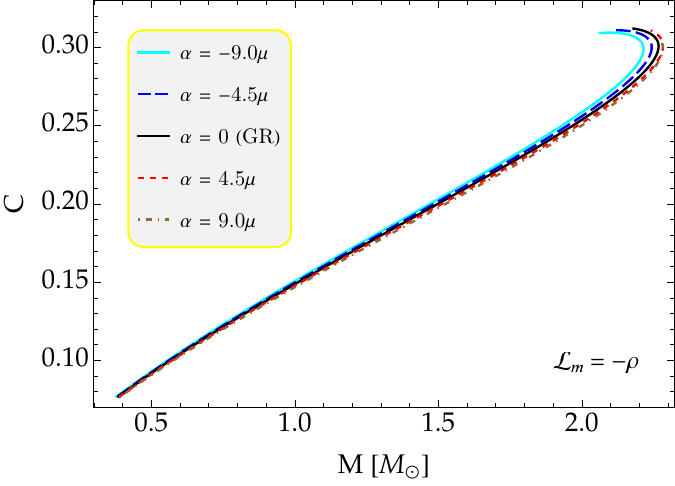}
    \includegraphics[width = 8.5cm]{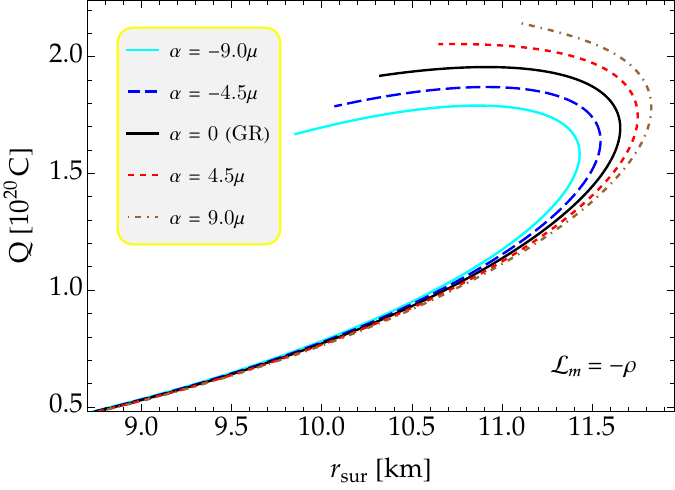}
    \caption{Mass-radius (top left), mass-central energy density (top right), compactness-mass (bottom left), and charge-radius (bottom right) diagrams for charged quark stars with MIT bag model EoS in the EMSG theory for the matter Lagrangian density $\mathcal{L}_m= -\rho$. Notice that we have used the parameter set: $B = 60\, \rm MeV/fm^3$, $\beta = 0.4$, and $\alpha \in [-9.0, 9.0] \mu$ where $\mu= 10^{-38}\, \rm cm^3/erg$. As in Fig.~\ref{FigsLmp_vary_alpha}, the black curves stand for the pure GR solutions in all plots.}
    \label{FigsLmRho_vary_alpha}
\end{figure}

\begin{figure}[h]
    \centering
    \includegraphics[width = 8.58cm]{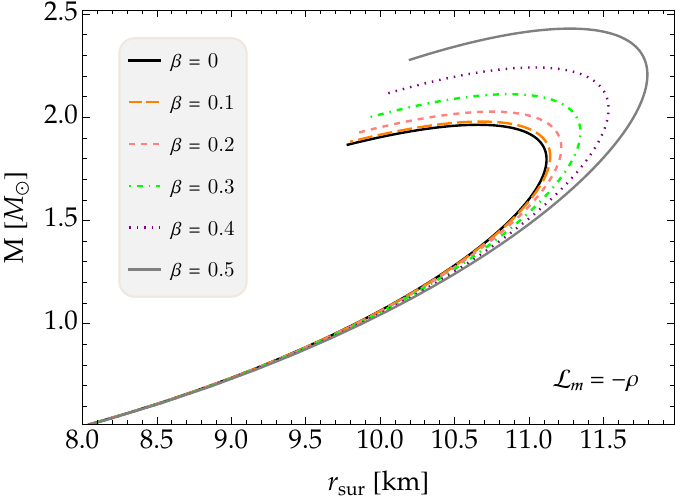}
    \includegraphics[width = 8.5cm]{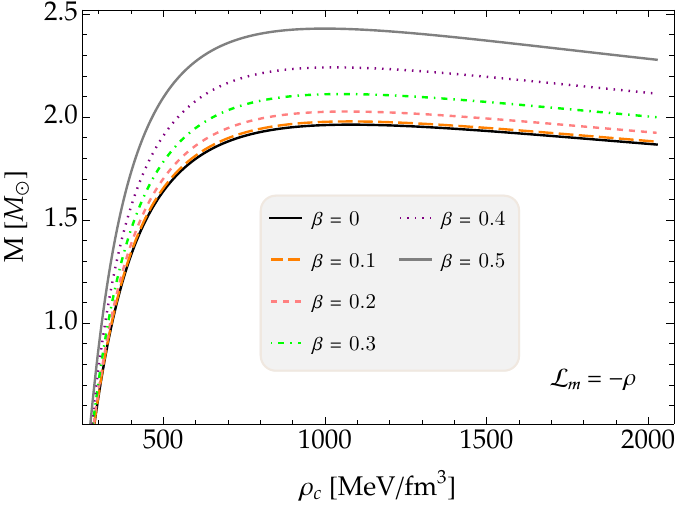}
    \includegraphics[width = 8.58cm]{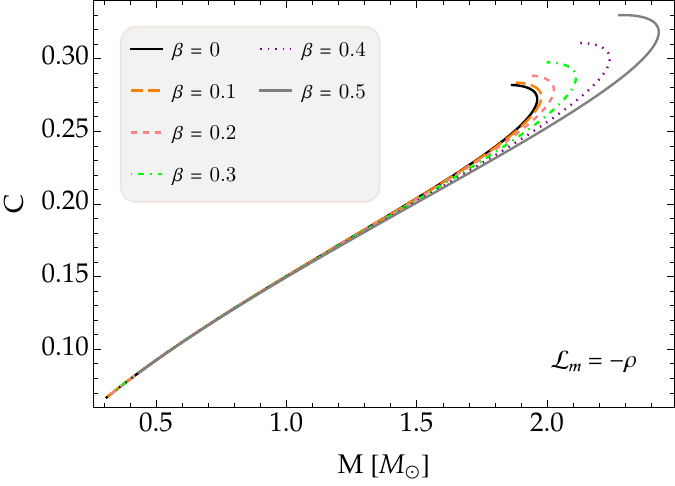}
    \includegraphics[width = 8.5cm]{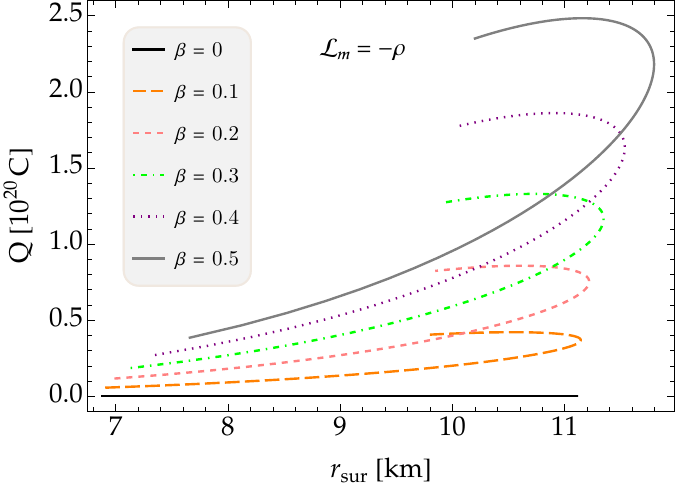}
    \caption{Macroscopic features of charged quark stars predicted by EMSG theory when $\alpha= -5.0\mu$ and $\mathcal{L}_m= -\rho$. We have also adopted $B = 60\, \rm MeV/fm^3$ for the bag constant and we have varied the charge parameter in the range $\beta \in [0, 0.5]$.}
    \label{FigsLmRho_vary_beta}
\end{figure}

In addition, considering the choice $\mathcal{L}_m= -\rho$, the sequence of charged equilibrium configurations is obtained by solving the stellar structure equations (\ref{TOV1LmMinusRho})-(\ref{TOV3LmMinusRho}) with boundary conditions (\ref{BC1}). The corresponding global characteristics associated with such configurations for fixed $\beta= 0.4$ and varying $\alpha$ are displayed in Fig.~\ref{FigsLmRho_vary_alpha}. Contrary to the other choice for the matter Lagrangian density, we see here that the largest values of maximum mass and maximum total charge are obtained when $\alpha >0$. Meanwhile, the numerical solutions for the case where $\beta$ varies and $\alpha$ is fixed, are displayed in Fig.~\ref{FigsLmRho_vary_beta}. We highlight that these results are qualitatively similar to those obtained in Fig.~\ref{FigsLmp_vary_beta} for the choice $\mathcal{L}_m= p$. The specific numerical values of the mass, radius, compactness and total charge for the maximum-mass configurations generated by both choices of $\mathcal{L}_m$ can be found in Tables \ref{tableVaryAlpha} and \ref{tableVaryBeta}.

\begin{table}[h]
    \caption{Global properties for maximum-mass configurations in energy-momentum squared gravity. The results are given for different values of $\alpha$ and fixed $\beta= 0.4$, where the gravity model parameter $\alpha$ is given in $\mu= 10^{-38}\, \rm cm^3/erg$ units. The energy-density values correspond to the critical central density where the function $M(\rho_c)$ is a maximum. The compactness $C= M/R$ is determined as a dimensionless quantity. Note that the particular value of $\alpha =0$ leads to the GR solution. The critical central density for each choice $\mathcal{L}_m= p$ and $\mathcal{L}_m= -\rho$ can also be found in the upper right plots of Figs.~\ref{FigsLmp_vary_alpha} and \ref{FigsLmRho_vary_alpha}, respectively.}
    \begin{ruledtabular}
    \begin{tabular}{c|cccccc}
    $\mathcal{L}_m$  &  $\alpha [\mu]$  &  $\rho_c [\rm GeV/fm^3]$  &  $r_{\rm sur} [\rm km]$  &  $M [M_{\odot}]$  &  $C$  &  $Q [10^{20} C]$ \\
    \hline
      &  $-3.0$  &  1.633  &   11.171  &  2.362  &  0.464  &  2.188 \\
      &  $-1.5$  &  1.238  &   11.191  &  2.309  &  0.453  &  2.043 \\
    $p$  &  $0$  &  1.105  &   11.141  &  2.268  &  0.447  &  1.946 \\
      &  $1.5$  &  1.027  &   11.077  &  2.231  &  0.442  &  1.867 \\
      &  $3.0$  &  0.973  &   11.009  &  2.198  &  0.439  &  1.798 \\
    \hline
      &  $-9.0$  &  0.992  &   10.947  &  2.215  &  0.445  &  1.788 \\
      &  $-4.5$  &  1.037  &   11.050  &  2.244  &  0.446  &  1.866 \\
    $-\rho$  &  $0$  &  1.105  &   11.141  &  2.268  &  0.447  &  1.946 \\
      &  $4.5$  &  1.222  &   11.212  &  2.283  &  0.447  &  2.030 \\
      &  $9.0$  &  1.539  &   11.248  &  2.283  &  0.446  &  2.117 \\
    \end{tabular}
    \end{ruledtabular}
    \label{tableVaryAlpha}
\end{table}

\begin{table}[h]
    \caption{Maximum-mass stellar configurations with MIT bag model EoS in EMSG for fixed $\alpha= 2.0\mu$ and varying the charge parameter $\beta$. The particular case $\beta= 0$ corresponds to the uncharged solutions. For the case $\mathcal{L}_m= p$, the maximum-mass values shown here can be visualized in the upper panels of Fig.~\ref{FigsLmp_vary_beta}. We observe that the critical central density is found to be a smaller and smaller value as $\beta$ increases. Furthermore, regardless of the choice for $\mathcal{L}_m$, the radius and mass of such configurations undergo a significant increase as a consequence of an increasing amount of electric charge. }
    \begin{ruledtabular}
    \begin{tabular}{c|cccccc}
    $\mathcal{L}_m$  &  $\beta$  &  $\rho_c [\rm GeV/fm^3]$  &  $r_{\rm sur} [\rm km]$  &  $M [M_{\odot}]$  &  $C$  &  $Q [10^{20} C]$ \\
    \hline
      &  0  &  1.063  &   10.643  &  1.935  &  0.399  &  0 \\
      &  0.1  &  1.059  &   10.667  &  1.951  &  0.402  &  0.414 \\
    $p$  &  0.2  &  1.049  &   10.741  &  2.000  &  0.409  &  0.845 \\
      &  0.3  &  1.032  &   10.867  &  2.086  &  0.422  &  1.313 \\
      &  0.4  &  1.007  &   11.055  &  2.220  &  0.441  &  1.843 \\
    \hline
      &  0  &  1.195  &   10.739  &  1.941  &  0.397  &  0 \\
      &  0.1  &  1.193  &   10.765  &  1.960  &  0.400  &  0.437 \\
    $-\rho$  &  0.2  &  1.185  &   10.842  &  2.017  &  0.409  &  0.896 \\
      &  0.3  &  1.171  &   10.977  &  2.118  &  0.424  &  1.400 \\
      &  0.4  &  1.148  &   11.176  &  2.276  &  0.447  &  1.983 \\
    \end{tabular}
    \end{ruledtabular}
    \label{tableVaryBeta}
\end{table}

A better way to quantify the deviations caused by the $\alpha T_{\mu\nu}T^{\mu\nu}$ term, with respect to the pure general relativistic case, on the maximum-mass values is by defining the following relative difference for a given value of $\alpha$:
\begin{equation}\label{RelativeEq}
    \Delta = \left. \frac{M_{\rm max, EMSG} - M_{\rm max, GR}}{M_{\rm max, GR}} \right\vert_\alpha ,
\end{equation}
where $M_{\rm max, GR}$ represents the maximum-mass value obtained in Einstein gravity (i.e., when $\alpha= 0$). Such deviations as functions of the charge parameter $\beta$ are shown in Fig.~\ref{FigsRelDeviation} for the two choices: $\mathcal{L}_m= p$ (left plot) and $\mathcal{L}_m= -\rho$ (right plot). In the case where $\mathcal{L}_m= p$ and $\alpha= -3.0\mu$, the maximum mass exhibits relative deviations of up to $\sim 8\%$ for charged QSs (with $\beta= 0.55$), while in the uncharged scenario ($\beta= 0$) the deviation grows to about $1.1\%$. 

Nevertheless, the behavior of these relative deviations as a function of $\beta$ is less trivial when $\mathcal{L}_m= -\rho$. For instance, when $\alpha= -9.0\mu$, the deviations are positive up to $\beta\sim 0.18$ and then turn negative. In other words, the maximum-mass values increase with respect to their GR counterpart until reaching a certain value of $\beta$ and then decrease. When $\beta= 0.6$, it is possible to reach deviations of up to $5.7\%$ for the gravity model parameter $\alpha= 6.0\mu$. Notably, for uncharged stars ($\beta =0$), when we increase $\alpha$ (from $-9.0\mu$) the values of $M_{\rm max}$ first increase and after reaching a maximum start to decrease. This behavior can also be observed in the right panel of Fig.~\ref{FigsUnchargedStars}.

\section{Concluding remarks }\label{sec5}

In this work we have analyzed the properties of specific stellar models for QSs in the recently proposed energy-momentum squared gravity (EMSG), where the free parameter of the theory $\alpha$ measures the deviations from GR. An important feature of this theory is the addition of nonlinear $T_{\mu \nu}T^{\mu \nu}$ term to the generic action. In a cosmological context, the EMSG adds novel features that could explain the recent cosmological observations. Here, our interest was to study the astrophysical consequences of Maxwell-EMSG theory on the relativistic structure of compact stars which are composed of strange quark matter. In particular, we derived the stellar structure equations for a spherically symmetric distribution of a charged perfect fluid assuming two different choices for the matter Lagrangian density ($\mathcal{L}_m= p$ versus $\mathcal{L}_m= -\rho$) within the framework of EMSG theory and we investigated their 
numerical solutions with proper boundary conditions as specified in Eq.~(\ref{BC1}). In addition, we have adopted a charge distribution where the electric charge density is proportional to the standard energy density, where emerges the charge parameter $\beta$ that measures the amount of charge in the stellar fluid.

In the gravitational context of the new theory we have investigated the impact of the $\alpha T_{\mu\nu}T^{\mu\nu}$ term on both uncharged ($\beta=0$) and charged ($\beta\neq 0$) quarks stars. Specifically, some important relations of charged spheres such as mass-radius, mass-central energy density, compactness-mass and charge-radius diagrams for QSs were analyzed. We summarize our findings step-by-step:
\begin{enumerate}
  \item In the first case, we examined the effect of the coupling constant $\alpha$ for uncharged QSs, i.e.~when $\beta= 0$. We separately discussed two different choices for the matter Lagrangian density in Fig.~\ref{FigsUnchargedStars}. For the choice of $\mathcal{L}_m = p$, we saw that the maximum gravitational mass and its corresponding radius increase (decrease) as $\alpha$ decreases (increases). Next, we extended the analysis for $\mathcal{L}_m = -\rho$. It could be observed that by decreasing $\alpha$, the maximum mass of QS increases and after reaching a maximum value it start to decrease. By comparing our results in the two approaches, we can say that $\mathcal{L}_m = p$ case is more effective to obtain maximum masses of QS depending on the value of $\alpha$. 
  
  \item We then explored QSs for $\mathcal{L}_m= p$ in the presence of electric charge, i.e.~when $\beta\neq 0$. We obtained the quark-star properties for available parameters when varying $\alpha$ and $\beta$, separately. For a fixed value of $\beta= 0.4$ and varying $\alpha$, we found that the mass-radius increase as $\alpha$ becomes more negative, while the situation is reversed for positive $\alpha$. For the maximum-mass configuration in the case of $\alpha= -3.0\mu$, $M_{\rm max}$ is found to be $2.362\, M_{\odot}$ with total charge $2.188 \times 10^{20}\, \rm Coulomb$, while in the pure GR scenario we get $M_{\rm max} = 2.268\, M_{\odot}$, see Fig.~\ref{FigsLmp_vary_alpha} and Table \ref{tableVaryAlpha}. Comparing the GR and EMSG results, we can say that EMSG gives higher masses compared with GR for negative values of $\alpha$.

  \item In the same spirit when we vary $\beta$ and keep $\alpha$ fixed, we observed that the maximum-mass values increase significantly as $\beta$ increases. The maximum value for the gravitational mass of a charged QS is $M_{\rm max} = 2.220\, M_{\odot}$ for $\beta= 0.4$ with total charge $1.843 \times 10^{20}\, \rm Coulomb$, see Table \ref{tableVaryBeta} when $\mathcal{L}_m= p$. In Fig.~\ref{FigsLmp_vary_beta}, we displayed our findings for the different basic properties of charged quark stars given a fixed value of $\alpha= 2.0\mu$ and varying $\beta$. Indeed, we saw that the total charge is strongly dependent on the charge parameter $\beta$, and therefore the total mass and total charge of the system increases with increasing $\beta$.

  \item Next, the results for QSs are obtained on the assumption of $\mathcal{L}_m= -\rho$ for fixed $\beta= 0.4$ and varying $\alpha$. One could observe the usual increasing of the maximum gravitational mass and goes up to $M_{\rm max} = 2.283\, M_{\odot}$ for $\alpha >0$. In this case, the total charge is found to be $2.117 \times 10^{20}\, \rm Coulomb$. Meanwhile, in conventional Einstein gravity, we recorded the maximum mass being $2.268\, M_{\odot}$ with total charge $1.946 \times 10^{20}\, \rm Coulomb$. Furthermore, these stars 
  have larger radii compared to the GR counterpart for $\alpha> 0$, see Fig.  \ref{FigsLmRho_vary_alpha} and Table \ref{tableVaryAlpha}.

  \item We also varied $\beta$ and observed that the obtained results are qualitatively similar to those obtained in Fig.~\ref{FigsLmp_vary_beta} for the choice $\mathcal{L}_m= p$. To be more specific, the maximum mass is $M_{\rm max} = 2.276\, M_{\odot}$ with total electric charge $1.983 \times 10^{20}\, \rm Coulomb$ within the framework of EMSG, see Fig.~\ref{FigsLmRho_vary_beta} and Table \ref{tableVaryBeta}.
 
  \item Finally, calculating the maximum-mass value for a given value of $\alpha$, we have quantified the relative deviations from the predictions of EMSG theory and its GR counterpart. For the case of $\mathcal{L}_m= p$, we observed that the maximum-mass deviation goes up to $\sim 8\%$ for charged QSs, while in the uncharged case the deviation grows to about $1.1\%$. Nevertheless, this relative deviation is less trivial when $\mathcal{L}_m= -\rho$ as observed from the right panel of Fig.~\ref{FigsRelDeviation}. Remarkably, the largest deviations occur for the charged case, which leads us to conclude that the parameter $\beta$ has a greater impact than $\alpha$ on the global properties of a compact star.
 
\end{enumerate}

Based on the obtained results, we can conclude that charged compact stars could existence in EMSG theory with maximum masses of around $2M_{\odot}$, as strange quark stars. In our study we have derived for the first time the modified TOV equations in EMSG not only for the usual choice $\mathcal{L}_m= p$, but also for $\mathcal{L}_m= -\rho$ and we have discussed in detail the effect of each one on the relativistic structure of quark stars in the presence of electric charge. It is worth mentioning that, the stellar structure equations obtained here could be applied to other more realistic equations of state.

\begin{acknowledgments}
JMZP acknowledges financial support from the PCI program of the Brazilian agency ``Conselho Nacional de Desenvolvimento Cient{\'i}fico e Tecnol{\'o}gico''--CNPq. T.~Tangphati is supported by School of Science, Walailak University, Thailand.
\end{acknowledgments}\

\end{document}